# The Effects of Object Shape, Fidelity, Color, and Luminance on Depth Perception in Handheld Mobile Augmented Reality

Tiffany D. Do [1], Joseph J. LaViola Jr. [2], Ryan P. McMahan [3]

Department of Computer Science, University of Central Florida, Orlando, FL, USA

**ABSTRACT**

Depth perception of objects can greatly affect a user's experience of an augmented reality (AR) application. Many AR applications require depth matching of real and virtual objects and have the possibility to be influenced by depth cues. Color and luminance are depth cues that have been traditionally studied in two-dimensional (2D) objects. However, there is little research investigating how the properties of three-dimensional (3D) virtual objects interact with color and luminance to affect depth perception, despite the substantial use of 3D objects in visual applications. In this paper, we present the results of a paired comparison experiment that investigates the effects of object shape, fidelity, color, and luminance on depth perception of 3D objects in handheld mobile AR. The results of our study indicate that bright colors are perceived as nearer than dark colors for a high-fidelity, simple 3D object, regardless of hue. Additionally, bright red is perceived as nearer than any other color. These effects were not observed for a low-fidelity version of the simple object or for a more-complex 3D object. High-fidelity objects had more perceptual differences than low-fidelity objects, indicating that fidelity interacts with color and luminance to affect depth perception. These findings reveal how the properties of 3D models influence the effects of color and luminance on depth perception in handheld mobile AR and can help developers select colors for their applications.

**Index Terms**: Human-centered computing—Human computer interaction (HCI)—HCI design and evaluation methods—User studies

## 1 INTRODUCTION

Color and luminance are well-studied depth cues that can affect what objects are perceived as nearer or further in a visual scene. This phenomenon has been documented by psychology and vision researchers dating back to the 19th century [2, 5] and artists such as Leonardo Da Vinci used these depth cues in his work [31].

Depth perception of virtual objects is an important factor to a user's experience in visual applications, especially in extended reality where users expect their actions in the virtual space to match the physical space [55]. Kruijff et al. [26] have described depth distortion as one of the most common problems in augmented reality (AR). For example, Swan et al. [53] observed that users had difficulty depth matching virtual objects. Singh et al. [47] have also conveyed the importance of accurate depth perception in AR and provided examples of AR applications that require depth matching between real and virtual objects. Similarly, Kalia et al. [24] remarked that inaccurate depth perception poses a problem in medical AR applications and causes difficulty for users. Hebborn et al. [22] have also expressed how inaccurate rendering of virtual objects often breaks the illusion of co-existence in AR applications, further providing compelling reasons to ensure that depth perception of virtual objects is accurate. Although most of these works have focused on head-mounted displays (HMDs), there has been recent interest in examining handheld mobile displays as well [4, 11, 12, 18]. In this paper, we use the term "handheld mobile AR" to refer to smartphone and tablet systems used to display virtual content over views of the real world provided by a back-facing camera [45].

Due to the importance of depth perception in AR, it is useful to investigate depth cues, such as color and luminance. The warmth and coolness of a color has a notable effect on depth perception. Several studies have indicated that warm colors, such as red, are perceived as nearer to a user while cool colors, such as blue, are perceived further away [5, 10, 20, 50]. Due to this effect, these colors have been respectively referred to as "advancing" colors and "retiring"/"retreating" colors by vision researchers [30, 37, 41]. Luminance is a well-studied depth cue and there is some evidence that suggests that it is a stronger depth cue than color hue [35, 38]. We refer to luminance as the relative brightness and darkness of a color, where a bright color has high luminance and a dark color has low luminance. Several studies have indicated that bright colors are perceived as nearer to a user, while dark colors are perceived as further [2, 8, 54]. This phenomenon occurs due to the luminance contrast between the object color and the background [34]. Multiple studies have indicated that a color with a higher luminance contrast to the background is perceived as further than a color with a lower luminance contrast [15, 21, 34].

While the effects of color and luminance as depth cues have been thoroughly studied, much of the work done has investigated real objects [2, 8, 37, 54]. Previous studies involving virtual stimuli have mostly focused on 2D simple shapes or lines [10, 13, 17, 19]. There are not many studies that investigate how the properties of 3D virtual objects interact with color and luminance to affect depth perception, especially in applications of augmented and mixed reality. Furthermore, studies using virtual 3D objects have typically evaluated a single 3D object [3, 47] and have not investigated the effects of object shape and fidelity on depth perception.

In this paper, we investigate the effects of object shape, fidelity, color, and luminance on depth perception of 3D objects in handheld mobile AR. Mobile handheld AR has become popular due to the ubiquitous nature of smartphones [45]. Advances in mobile technology and affordability have allowed mobile AR to become increasingly common in a variety of fields [7, 40]. Evaluating depth perception in mobile handheld AR will provide useful and ecologically valid results that can help developers choose colors in applications.

In order to evaluate depth perception, we conducted a preliminary paired comparison experiment to group colors into distinct groups of similar perception. We administered this procedure for objects of varying shape and fidelity. The results of our study indicate that there is some interaction between properties of 3D models and the effects of color and luminance on depth perception. For a simple, high-fidelity 3D shape, bright colors were perceived as nearer than dark colors, regardless of hue. Additionally, red was perceived as nearer than any other color, but this only occurred when the luminance was high. This effect was not observed for a low-fidelity version of the simple shape or for a complex 3D shape. There does not seem to be an effect of color on complex 3D shapes. However, there appears to be some effect of luminance on complex 3D shapes, as a color will generally appear closer than a darker version of the same color, regardless of fidelity.

---
[1] tiffanydo@Knights.ucf.edu
[2] jjl@eecs.ucf.edu
[3] rpm@ucf.edu

## 2 RELATED WORK

### 2.1 Depth Perception in AR and VR

Depth perception has been a widely studied topic in AR and virtual reality (VR), but most work has focused on HMDs [4]. Swan et al. [51] provided an excellent overview of previous work of depth perception in AR and VR and noted that distances are consistently underestimated in VR scenes. Thompson et al. [55] proposed that image quality was not the reason for underestimation in VR after investigating the effects of environment quality on depth perception. Interestingly, Kjelldahl and Prime [25] noted that 3D wireframe objects were more difficult for participants to make depth judgments on compared to a solid object on a desktop monitor. These studies are highly relevant to our experiment as investigations of the effects of object fidelity on depth perception. However, they are focused on different platforms and do not investigate those effects for handheld mobile AR devices.

Swan et al. [51] also surveyed several methods for estimating depth in mixed reality, such as perceptual matching, blind walking, triangulation by walking, and forced-choice tasks. They described a forced choice task that requires observers to make discrete depth choices, such as whether one object is nearer or further than another [51]. We chose to use a forced-choice task, as opposed to a perceptual matching task, due to the simplicity of forced-choice tasks. Perceptual matching is a more complicated task that requires users to adjust the position of a target object until it appears to be the same depth as a reference object. Due to the outbreak of COVID-19, participants completed our study remotely using their own mobile device and were unable to interact face to face with the experimenter, thus necessitating a simple task. Additionally, forced-choice tasks have seen success in analyzing color as a depth cue on objects displayed on a desktop monitor [3].

Depth perception in AR has seen notable work in the last decade. Swan et al. [53] investigated depth matching in AR and found that observers overestimated the matching distance of a virtual object compared to a real object at a reaching distance, although this study was conducted using head-worn AR. Rosales et al. [44] also found that distances to off-ground objects were perceived differently than distances to on-ground objects in a study using head-worn AR. Kruijff et al. [26] provided a thorough overview of perception issues in a wide range of AR platforms. They described depth distortion as one of the most common issues in AR and noted that relative brightness and color can cause problems in depth perception in both handheld mobile AR and head-worn AR. They also described other pictorial depth cues, such as occlusion, relative size, and aerial perspective [26]. Adams [1] noted that distance perception in AR is inconsistent and that consistent depth cues may improve depth perception in AR. Adams focused on shadows as depth cues, while our study focuses on color and luminance.

More recently, there has been interest in examining depth perception in handheld mobile AR. Swan et al. [52] were the first to investigate distance judgments of real and virtual objects in a tablet AR system. They found that a user's picture perception in tablet-based AR devices is fundamentally different than HMD-based VR and AR. Dey et al. [11] studied the effect of resolution and display size on depth perception in handheld mobile AR and found that resolution had no significant effect, but that a smaller display size caused less underestimation. Participants in this experiment used either a mobile phone or tablet [11]. Liu et al. [29] investigated depth perception of virtual objects in handheld augmented AR and noted that virtual objects were perceived as further than real object. Berning et al. [4] investigated the effects of stereoscopic handheld mobile AR on depth perception and proposed that depth judgment is mostly influenced by monoscopic depth cues. Although they did not study color and luminance as a depth cue in AR, they noted that these cues may have influenced their results and should be studied further [4]. Our study directly addresses this issue to determine if color and luminance have effects on depth perception in handheld mobile AR systems.

### 2.2 Color Hue and Luminance as Depth Cues

Most previous experimental work on color hue and luminance as depth cues was done in the mid-19$^{th}$ century. These studies largely followed a similar consensus: warm colors are perceived as nearer than cool colors and bright colors are perceived as nearer than dark colors [8, 20, 23, 30, 33, 37, 38, 54, 57]. Multiple studies have indicated that the difference in luminance contrast between an object and the background causes this effect, and the majority of these studies used dark backgrounds [15, 21, 34]. Many of these early studies examined simple 2D stimuli, such as shapes or letters, but the effects of color hue and luminance on depth perception have also been noted in real 3D objects [23, 33].

There exists some evidence that the effects of color on depth perception can be reversed depending on luminance. Pillsbury and Schaefer [41] proposed that blue light is perceived as closer than red light. Olgunturk [36] argued that this phenomenon is due to the Purkinje shift, where blue light appears brighter than red light at low luminance levels. Both Olgunturk [36] and Payne [39] have proposed that luminance is a stronger depth cue than color, which explains the results of Pillsbury and Schaefer [42].

Table 1: Overview of previous studies involving color hue or luminance as depth cues for virtual objects.

|  |  | Virtual Object | | | Color | |
| --- | --- | --- | --- | --- | --- | --- |
| Study | Platform | Dimensions | Shape | Fidelity | Hue | Luminance |
| [10] | Desktop | 2D | ✓ | ✗ | ✓ | ✗ |
| [13] | Desktop | 2D | ✓ | ✗ | ✗ | ✓ |
| [19] | Desktop | 2D | ✗ | ✗ | ✓ | ✓ |
| [17] | Desktop | 2D | ✗ | ✗ | ✗ | ✓ |
| [3] | Desktop | 2D/3D | ✓ | ✗ | ✓ | ✗ |
| [24] | Desktop | 3D | ✗ | ✗ | ✓ | ✗ |
| [6] | VR Screen | 3D | ✗ | ✗ | ✗ | ✓ |
| [47] | AR Haploscope | 3D | ✗ | ✗ | ✗ | ✓ |
| **Ours** | **AR Handheld** | **3D** | ✓ | ✓ | ✓ | ✓ |

However, these early experiments studied the effect of color and luminance as depth cues in real objects, while we are interested in virtual objects. More recent studies have investigated depth perception of virtual objects. In Table 1, we provide a brief overview of these studies. Most of these studies used a desktop monitor and did not focus on the properties of 3D virtual objects, such as shape and fidelity. Additionally, we explore both color hue and luminance as depth cues.

One of the earliest studies using virtual stimuli was executed by Dengler and Nitschke [10]. They investigated the effects of background luminance on the traditional warm-cool depth effect. In an experiment involving simple 2D lines and boxes, they found that warm colors appeared closer on a black background, but cool colors appeared closer on a white background [10]. This effect was consistent with earlier non-virtual observations by Hartridge [20].

Some studies have explored the interaction of multiple depth cues using simple 2D stimuli. Dresp et al. [13] proposed that the luminance contrast of an object against the scene is a principal pictorial cue, even when there are other depth cues in the scene, such as occlusion and interposition. Objects with a higher luminance contrast were more likely to be selected as near, regardless of other depth cues [13]. Guibal and Dresp [19] later found that color is not an independent depth cue and is strongly influenced by luminance contrast and stimulus geometry. Fujimara and Morishita [17] explored the combination of saturation and brightness as depth cues and found that the combination of the two cues was a stronger depth cue than brightness or saturation alone [17]. Although these studies provide evidence that depth cues can have interactions, they are only focused on 2D simple stimuli, such as shapes and lines, whereas we focus on 3D objects.

The work of Bailey et al. [3] is among the most relevant for our experiment. They studied the effect of color on depth perception of a realistic 3D object in comparison to a traditional stimulus of a 2D simple shape. It was found that warm colors appeared closer than cool colors for the 2D simple shape, but this effect was not strong for the realistic 3D object. The colors tested in the experiment were of equal luminance with low saturation. Bailey et al. [3] provided compelling evidence that model dimension or shape may interact with color as a depth cue, but only studied one 3D model, while we investigate the properties of multiple 3D models.

Investigating color and luminance as depth cues in AR has received interest in recent years. Singh et al. [47] examined brightness of a virtual object as a depth cue in AR using a custom-built AR haploscope. They found that the brightness of a virtual object in AR has effects on depth perception and depth matching. This experiment provided useful insights on brightness as a depth cue for head-worn AR, but we are interested in handheld mobile AR. Kalia et al. [24] expressed motivation to improve depth perception in AR medical imaging applications. They combined color cues with depth of focus blur and found that using warm colors for close objects and cool colors for far objects improved perception. Although this study was motivated by AR applications, they did not use an AR apparatus and showed participants images or videos on a desktop monitor [24].

Castell et al. [6] investigated how the brightness of a ceiling in VR affected perception of the ceiling height and found that a brighter ceiling caused participants to perceive the ceiling as higher than a darker ceiling. Castell et al. [6] noted that this result followed traditional interior design theory, where brighter ceilings are perceived as higher than darker ceilings. This study focused on the brightness of the interior space instead of a 3D object in the scene, which our study addresses.

## 3 HANDHELD MOBILE AR DEPTH PERCEPTION STUDY

We conducted a paired comparison experiment to determine the effects of object shape, fidelity, color, and luminance on depth perception of 3D objects in handheld mobile AR. We expected simple shapes to be more affected by color and luminance than complex shapes and high-fidelity objects to be more affected than low-fidelity objects. We expected warm colors to still be perceived as nearer than cool colors of the same luminance, and we expected that brighter colors would be perceived nearer than darker colors, regardless of shape, fidelity, and color hue.

### 3.1 Paired Comparison Method

The method of paired comparisons is a well-regarded scaling method that places objects along a continuum of quality [14]. An object's placement depends on the degree that it exhibits some common property, which can be any qualitative or quantitative attribute that can be compared between objects [48, 56]. This method presents two objects to a participant and asks them to select the object that more exhibits the tested property. This procedure is then repeated for all pairs of objects [48].

The paired comparison method was first introduced by Thurstone in 1927 [56] and has since been expanded on several times [48, 49]. It has been used in many experiments that analyze visual stimuli with some examples being [3, 27, 28, 43, 58]. It has also been used to analyze depth perception in particular [3, 28].

We decided on using the paired comparison method due to the nature of our experiment. Ledda et al. [27] presented strong reasoning for using this method. Mantiuk et al. [32] compared multiple subjective methods (single-stimulus, double-stimulus, forced-choice pairwise comparison, and similarity judgments) for image quality assessment and concluded that the pairwise comparison method was the most accurate, time efficient, and least subject to measurement variance.

### 3.2 Experimental Design

We conducted a 2 x 2 x 6 x 2 within-subject experiment to evaluate the effects of object shape (2 shapes), fidelity (2 levels), color (6 hues), and luminance (2 levels) on depth perception for handheld mobile AR applications.

#### 3.2.1 Object Shape and Fidelity

Table 2: The object shape and fidelity conditions used in our study.

|       |         | Fidelity       |                 |
|-------|---------|----------------|-----------------|
|       |         | Low            | High            |
| Shape | Simple  | Cube           | Sphere          |
|       | Complex | Low-poly Bunny | High-poly Bunny |

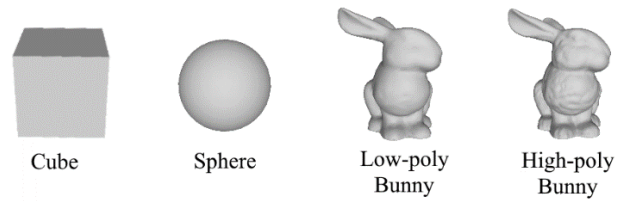

Figure 1: The models used in our paired comparisons study.

We investigated the effects of object shape at two levels: a simple sphere and the complex Stanford bunny. For the simple sphere, we used a high-poly sphere (768 triangles) for the high-fidelity condition and a cube (12 triangles), which is the lowest polyhedral approximation of a sphere [16], for the low-fidelity version of the sphere. For the complex bunny, we used the original high-poly version (69,630 triangles) for the high-fidelity condition and a reduced low-poly version (4,968 triangles) for the low-fidelity condition. Table 2 provides a summary of the four object conditions used in our study, and Figure 1 depicts the corresponding models.

Every participant completed four paired comparison tasks, one for each model, involving color hue and luminance, as described in the next section. To avoid potential ordering effects, such as learning or boredom, we counterbalanced the presentation order of the four models between subjects using a Latin squares design.

### 3.2.2 Color Hue and Luminance

We selected six color hues representing the spectrum of colors. We chose three warm colors (red, magenta, and yellow) and three cool colors (green, blue, and cyan). We included both a bright and dark version of each color, giving us a total of 12 color conditions. All colors were luminance-balanced for both bright and dark versions. All objects were virtually illuminated with a white (#FFFFFF) directional light that had an intensity value of 1.0. We used the same AR target background for each object so that the differences in luminance conditions created contrast.

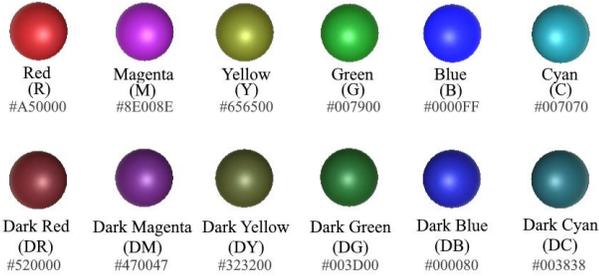

Figure 2: The color conditions used for the paired comparisons. A color is displayed with its given abbreviation and hex color code.

### 3.3 Task Evaluation

Figure 3: Example preference matrix for one participant when shown 12 color conditions of a given model. Each color in a row is compared with all other colors to establish a preference score.

Paired comparison experiments require participants to select between two objects based on a shared quality, as explained in Section 3.1. We presented the participant with two differently colored versions of the same virtual object via a handheld mobile AR interface and asked them to select the object that appeared closer to them. The participants were instructed to evaluate all possible comparison pairs from the set of color conditions.

Suppose that $n$ is the number of colors that we wish to compare against one another. For a given 3D model, each participant is presented with $(n(n-1))/2$ pairs. In our experiment, in which we have 12 color conditions, each participant compares 66 pairs of stimuli per model (see Figure 3). A user's vote is recorded for every selection. After evaluation of all pairs for a model, these votes are combined into a single $n \times n$ preference matrix. An example of one such preference matrix can be seen in Figure 3. If a matrix cell at row $i$ and column $j$ contains a 1, then this indicates that the participant selected the color at row $i$ as closer in depth than the color at row $j$. This also implies that the matrix cell at row $j$ and column $i$ would be filled with a 0. The preference matrices of all participants are then added together to determine overall scores of each color for an experimental condition. The participant repeats this process for all 4 models, leading to a total of 264 comparisons.

### 3.4 Research Questions and Hypotheses

**RQ1: How does the shape of a 3D model interact with color and luminance as depth cues?**

*H1 (Shape): Simple shapes will be more affected by color and luminance as depth cues than complex shapes.* Previous work found that a realistic 3D object was less subject to the effects of color as a depth cue than a 2D square, indicating a relationship between complexity and depth cues [3]. We expect that simple shapes will demonstrate classical effects of color and luminance as depth cues (i.e., warm colors are closer than cool colors, and bright colors are closer than dark colors), while complex shapes may demonstrate these effects to a lesser degree.

**RQ2: How does the fidelity of a 3D model interact with luminance and color as depth cues?**

*H2 (Fidelity): High-fidelity objects will be more affected by color and luminance as depth cues than low-fidelity objects.* Because related work indicated that a high-fidelity 3D object was easier to judge depth than a low-fidelity object [25], we expect that high-fidelity models will demonstrate classical effects of color and luminance as depth cues, while low-fidelity models may demonstrate these effects to a lesser degree.

**RQ3: How does the color of a 3D model affect its perceived depth?**

*H3 (Color): Warm colors will be perceived as nearer than cool colors of the same luminance.* We expect that warm bright colors will be perceived as nearer than cool bright colors, while warm dark colors will be perceived as nearer than cool dark colors. Previous studies propose that luminance is a stronger depth cue than color [35, 38], so we expect that this phenomenon will only occur at the same luminance level.

**RQ4: How does the luminance of a 3D model affect its perceived depth?**

*H4 (Luminance): Brighter colors will be perceived as nearer than darker colors, regardless of shape, fidelity, and color hue.* We expect that brighter colors will be perceived as nearer than dark colors, as they have more contrast with the background color.

### 3.5 Statistical Analysis

We employed the same methodology and analysis methods as Ledda et al. [27], who provided compelling reasons to use the following methods of analysis in a paired comparison experiment.

#### 3.5.1 Kendall Coefficient of Agreement

If all participants vote the same way, then there is complete agreement. However, this is rarely the case, and it is important to determine if there is actual agreement between participants. Kendall's coefficient of agreement utilizes the number of agreements between pairs.

$$\Sigma = \sum_{i \neq j} \binom{p_{ij}}{2} \quad (1)$$

In Equation 1, suppose that $p_{ij}$ is the number of times that $color_i$ is preferred to $color_j$. $\Sigma$ denotes the sum of the number of agreements in pairs, extending over all pairs excluding the diagonal because a color is never compared to itself. $\Sigma$ can then be used to calculate a coefficient of agreement among participants as defined by Kendall and Babington [48]:

$$u = \frac{2\Sigma}{\binom{s}{2}\binom{n}{2}} - 1 \quad (2)$$

Suppose that $n$ denotes the number of colors, while $s$ denotes the number of participants. If all participants made identical choices, then $u$ would be equal to 1. $u$ decreases as participants disagree, tending to $-1/(s-1)$ if $s$ is even and $-1/s$ if $s$ is odd. The coefficient of agreement $u$ can tell us how much the participants agree with one another. We can test the significance of $u$ to determine if participants actually agree with one another using a large sample approximation to the sampling distribution [46]:

$$\chi^2 = \frac{n(n-1)(1 + u(s-1))}{2} \quad (3)$$

$\chi^2$ is asymptotically distributed with $n(n-1)/2$ degrees of freedom. We can use a table of probability value for $\chi^2$, for example Table C in [46]. Using this statistic, we can test the null hypothesis $H_0$ that there is no agreement among participants, which implies that all colors are perceptually equivalent.

### 3.5.2 Coefficient of Consistency

Paired comparison experiments often measure the consistency, or transitivity, of a participant's choices. For example, if a participant selects that $color_A$ is closer than $color_B$ and that $color_B$ is closer than $color_C$, then they should also select that $color_A$ is closer than $color_C$. If they select otherwise, then they would have created a circular triad. Inconsistency can frequently occur when the items being compared are similar and it is difficult to make judgments. We calculated a coefficient of consistency $\zeta$ as defined by Kendall and Babington [48] for even $n$, where $c$ denotes the number of circular triads:

$$\zeta = 1 - \frac{24c}{n^3 - 4n} \quad (4)$$

We can determine the number of circular triads using the following formula [9]:

$$c = \frac{n}{24}(n^2 - 1) - \frac{1}{2}\sum(p_i - (n-1)/2)^2 \quad (5)$$

where $n$ is defined as the number of colors and $p_i$ is the score of each color. It is important to note that participants can have a high coefficient of agreement $u$ while having a low coefficient of consistency $\zeta$. Participants can individually make inconsistent decisions (i.e., make circular triads) and still have a high $u$ if they overall agreed on these inconsistent decisions. $\zeta \in [0,1]$, where a coefficient of 1 indicates perfect consistency. $\zeta$ tends towards 0 as inconsistency increases. We can use $\zeta$ to learn useful information about the similarity of colors being tested. We can expect a low value if colors are perceptually similar, which makes selection a difficult task.

### 3.6 Apparatus

Due to COVID-19, participants were instructed to use their own Android device. In 0, we list all the screen display size and resolution of all devices and number of users. All devices are recent phones released within the last five years and are running at least Android 4.4 or above. The AR application was developed using the Unity game engine and the Vuforia AR engine.

Table 3: Overview of specifications of devices used in our study.

| Display Size | Screen Resolution | Devices |
|---|---|---|
| 5.0 in. | 1280 x 720 | 2 |
| | 1920 x 1080 | 1 |
| 5.1 in. | 1440 x 2560 | 3 |
| 5.5 in. | 1080 x 2160 | 1 |
| 5.6 in. | 2220 x 1080 | 1 |
| 6.2 in. | 1080 x 2960 | 2 |
| | 1440 x 2960 | 1 |
| | 2160 x 1080 | 1 |
| 6.3 in. | 2380 x 1080 | 1 |
| 6.4 in. | 720 x 1560 | 1 |
| | 2960 x 1440 | 1 |
| 6.41 in. | 1080 x 2340 | 1 |

Participants were only allowed to use an Android mobile phone. Tablets were not used in the experiment due to the differences observed in depth perception between mobile phones and tablets in handheld AR [11]. Although screen resolution varies substantially across the devices, Dey et al. [11] noted that screen resolution did not affect depth perception in handheld AR. Participants were asked to print an 8.5 in. x 11 in. sheet of paper that contained two image targets. Two versions of the same virtual object were placed above the image targets for pairwise comparisons. The background of the paper was a neutral grey. The application set screen brightness to max and controlled the angle of the objects to the camera (i.e., no object is angled towards the camera more than the other). Participants were asked to position the image targets in the center of the screen at the beginning of each condition and were required to recalibrate if they moved away by more than 5% of the screen. Minor movement was permitted, and virtual objects had motion parallax consistent with the real world. The sizes of the virtual objects were controlled by the application and kept equal.

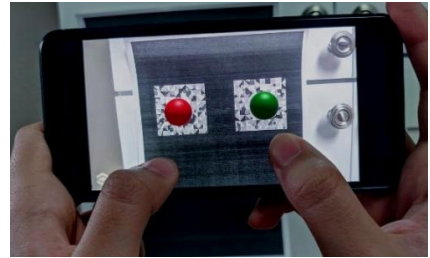

Figure 4: Depiction of our study's setup and pair comparisons task.

### 3.7 Procedure

Before participation in the study, participants filled out a background survey. They were then asked to print out the sheet of paper containing the image targets. Afterwards, participants installed an Android application on their device and were asked to disable all color distortion applications (e.g., blue light filter).

The participant would then undergo the paired comparison experiment with the first model in their assigned Latin squares ordering. The participant would be presented with a pair of objects respectively on the left and right side of the screen, each with a different color condition. They were instructed to select the object that appears closer to them based on their initial impulse. Selection was done by tapping on the object with their finger. After the result is recorded, the participant was presented with the next pair of color conditions after a 500-millisecond delay. This delay is based off of a similar successful paired comparison experiment that also evaluated depth perception [59]. Both the ordering of the pairs shown and the positions (left or right) of the color conditions within the pair was randomized to prevent any bias. This procedure was outlined by Dunn-Rankin et al. [14]. After all pairs of color conditions were presented to the participant, they were given the next object model in their assigned ordering. This process was repeated for all four models.

### 3.8 Participants

We recruited 16 unpaid volunteers (8 females, 8 males). Their mean age was 24.76 years, within a range from 21 to 35. Based on self-reported background data, 13 of the participants played video games with 3D models on a regular basis (i.e., at least one hour per week). All participants reported having normal or corrected vision and no color deficiencies (e.g., color blindness). Each participant was assigned to one of four ordering cohorts to counterbalance presentation order of the four models.

## 4 RESULTS AND DISCUSSION

### 4.1 Results

For each model, the preference matrices of all participants were tabulated in a combined preference matrix. This combined preference matrix adds the scores of all participants together. In Figure 5, we show the total scores of each color in every model. This score indicates how many times a color was selected over another color.

0 shows the statistical tests of all models. Although the average coefficient of consistency for each model is not very high, this is to be expected because several colors have similar perception and cannot be told apart. For example, the dark colors for most models generally have similar scores and it is likely that they are similar in terms of perception. We used a forced-choice pairwise comparison design without the possibility of ties, which may have caused participants to create circular triads when presented with perceptually similar colors. The average coefficient of consistency can only tell us how reliably individual colors can be ranked. We analyzed the average number of circular triads using the overall circularity test as described in [14] and rejected the null hypothesis $H_0$ that participants have the same circularity as choosing by chance with $\alpha = 0.05$ level and 11 degrees of freedom for all four models. Thus, we can conclude that participants do have some consistency.

Table 4: Overview of statistical analysis for all models.

|  | *Coeff Cons (ave) ζ* | *Coeff Agr u* | $\chi^2$ | *Significance p, 66 df* |
|---|---|---|---|---|
| Cube | 0.311 | 0.003 | 68.5 | > 0.1 |
| Sphere | 0.425 | 0.095 | 160.50 | < 0.001 |
| Low-poly Bunny | 0.305 | 0.021 | 86.5 | < 0.05 |
| High-poly Bunny | 0.373 | 0.091 | 156.26 | < 0.001 |

We analyzed the significance of the coefficient of agreement *u* for all models using an approximation as described in Section 3.5.1. If the p-value of the coefficient of agreement *u* is significant, then we can create groups of perceptual similarity, where all colors in a group are not perceived significantly differently. For the sphere, low-poly bunny, and high-poly bunny, we can reject the null hypothesis $H_0$ at $\alpha = 0.05$ level for 66 degrees of freedom to conclude that there is some agreement amongst participants. The p-value of the cube's coefficient of agreement was not significant, so we cannot conclude that there is agreement between participants, and therefore cannot create groups of perceptual similarity or state that any colors are perceptually different. A possible reason is that the cube is less susceptible to color and luminance as depth cues due to the sharp edges and shadows.

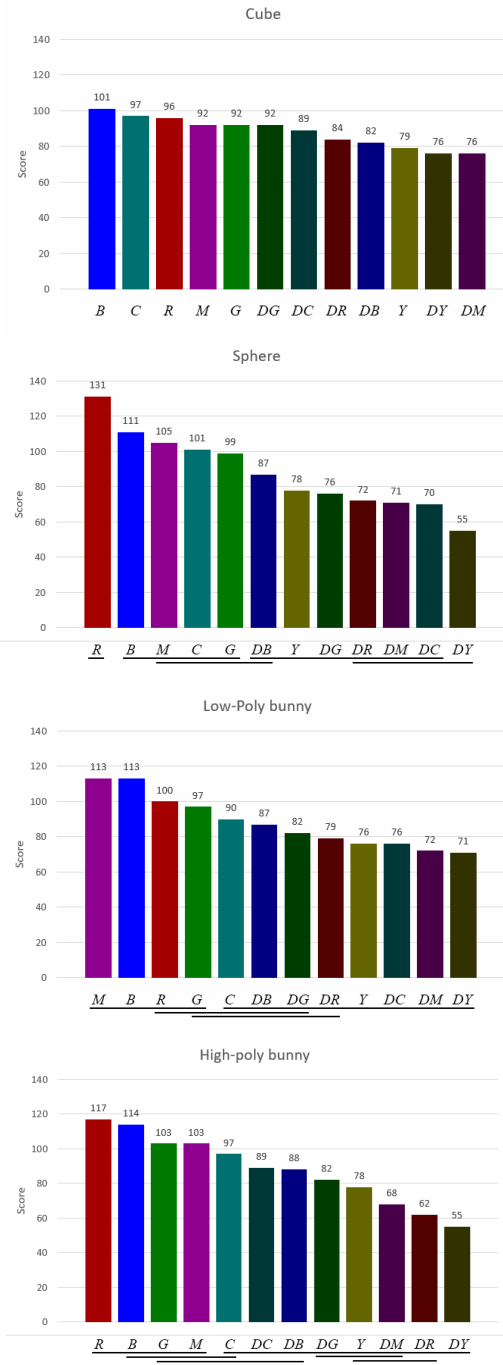

Figure 5: Bar graphs of total scores of each color condition per model. Colors are ordered from greatest score to least score. Perceptually similar groupings of models are presented, excluding the cube. Any colors whose scores are underlined are similar.

Table 5: Overview of total number of groupings and the largest number of distinct, non-overlapping groupings.

|  | Total Groupings | Distinct Groupings |
|---|---|---|
| **Cube** | 1 | 1 |
| **Sphere** | 5 | 3 |
| **Low-poly bunny** | 4 | 2 |
| **High-poly bunny** | 7 | 3 |

The cube was unable to be evaluated because the p-value of the coefficient of agreement *u* was not significant. Therefore, we cannot state that there is any difference in depth perception between color conditions for the cube. The groupings of colors for the sphere, low-poly bunny, and high-poly bunny are shown in Figure 5. We created groupings using the least significant difference method outlined by Starks and David [49], at the $\alpha = 0.05$ level. Table 5 shows the total number of groupings and largest number of distinct, non-overlapping groupings for each model. Note that the cube only has 1 grouping total as we cannot conclude that there is any difference in depth perception between color conditions.

We evaluated our hypotheses based on the total number of perceptual groupings, the largest number of distinct groupings, and pairs of stimuli that do not share a grouping. It is interesting to note that yellow was the only bright color that is consistently ranked lower than dark colors. No other bright color ranked below dark colors in terms of score. This is likely because the bright colors had a relatively low luminance due to luminance balancing, causing the yellow to appear more like brown. Bailey et al. [3] observed similar results in their study, where they used an unsaturated yellow with relatively low luminance. Due to this reason, we will mostly exclude it from our discussion.

Our results partially support H1 (Shape). We cannot conclude that the depth perception of the low-fidelity, simple shape (the cube) is affected by color and luminance. However, the high-fidelity, simple shape (the sphere) was the only model that demonstrated classical effects of color on depth perception. Red, the warmest color, is ranked as the closest color and is in a distinct group of its own. Red is perceived as significantly distinct from any other color, which was not the case for either bunny model.

The perceived depth of the sphere is also more affected by luminance compared to the complex bunny shapes. None of the bright colors are in a perceptually similar grouping as its darker counterpart. The complex shapes have many groupings that include both dark and bright colors, unlike the sphere. H1 (Shape) is supported for the high-fidelity objects. The depth perception of complex shapes may not be affected by color and luminance as much as a simple shape due to the presence of other depth cues such as shadows and contours. The bunny has multiple grooves and curves in the mesh, which give the viewer more shadows and highlights, causing them to rely less on color and luminance.

Our results support H2 (Fidelity). Increased fidelity increased perceptual differences among the color and luminance conditions for both shapes, as indicated by the number of groupings. The low-fidelity cube had only 1 total grouping, while the high-fidelity sphere had 5 total groupings and 3 distinct non-overlapping groupings. The low-fidelity bunny had 4 total groupings and 2 distinct groupings, while the high-fidelity bunny had 7 total groupings and 3 distinct groupings. A higher number of groupings indicates that participants observed more perceptual differences. Furthermore, the overall scores for red notably increased from the low fidelity to the high-fidelity versions of both shapes, showing that fidelity may have some effect on depth perception. Additionally, the overall scores for dark yellow notably decreased from the low fidelity to the high-fidelity versions of both shapes.

H3 (Color) was not supported by our results. Warm colors of the same luminance were not perceived as closer than cool colors for all objects, except for red in the sphere condition.

H4 (Luminance) was partially supported by our results. Except for the cube, most of the bright colors were perceived as significantly closer than their darker counterparts. For the sphere, all of the bright colors were perceived as closer. For the low-poly bunny, magenta, blue, and red were perceived closer than their darker counterparts, but green, cyan, and yellow were not. For the high-poly bunny, red, blue, green, and magenta were perceived closer than their darker counterparts. Only cyan and yellow were not perceived as significantly closer than their counterparts. Even for the cube, which had only one total grouping, the brighter colors have higher scores than their darker counterparts. Our relatively low sample size is the most likely cause for the lack of significant differences among all bright and dark color counterparts.

## 4.2 Discussion

Our preliminary results indicate that the shape and fidelity of 3D virtual objects interact with color and luminance to affect depth perception in handheld mobile AR. Color and luminance are well-studied depth cues, but the influence of these depth cues can vary depending on the shape and fidelity of a 3D object. A very warm color, such as red, affects the depth perception of a high-fidelity, simple shape, but only when the color is bright. The warmth and coolness of a color did not affect a complex shape, supporting results found by Bailey et al. [3] where the perceptual depth of a realistic 3D virtual object was not affected by color. High-fidelity objects had more perceptual differences than low-fidelity objects, indicating that fidelity interacts with color and luminance as depth cues.

We could not determine if the depth perception of the low-fidelity, simple 3D object (the cube) is affected by both color and luminance. The structure of the cube presents strong linear and angular information that is not available in the smooth, continuous curvature of the sphere, which likely explains the lack of significant differences for the cube. For all other models, all bright colors had higher total scores than dark colors, excluding yellow. We used the same background for all stimuli to create luminance contrast, which may have caused the bright colors to have higher scores.

AR developers should choose colors with caution if depth perception is important to the user experience. For example, if a developer does not want a simple object to appear closer than another simple object, they should ensure that the luminance of both objects is equal and that the color is not bright and warm. These findings can also be used to improve applications if distinct depth perception of objects is required. Kalia et al. [24] implemented an excellent example in which they used warm colors to represent closer objects and cool colors to represent further objects in medical images displayed on a desktop monitor. Our findings could help improve such an application where developers want to further emphasize depth using depth cues. If developers understand the effects of color and luminance on 3D objects, they can properly utilize these cues to their advantage.

## 4.3 Limitations

Due to the outbreak of COVID-19, we were unable to run a laboratory study. Users participated in this study remotely using their own mobile devices. There are differences among the devices, such as max screen brightness and screen color accuracy, that may have affected our results. Additionally, our sample size was relatively low, which likely decreased the number of potential color groupings. However, we believe that these preliminary results are ecologically valid and could possibly represent the effect of object shape, fidelity, color, and luminance on depth perception in the real world. Additionally, our user study only had 16 subjects and more subjects would probably provide more information and potentially more differences among the conditions.

## 5 CONCLUSION AND FUTURE WORK

In this paper, we presented a comparison of the effects of object shape, fidelity, color, and luminance on depth perception of 3D objects in handheld mobile AR. To determine these effects, we conducted a paired comparison experiment. The results of our study indicates that the shape and fidelity of 3D virtual objects interact with color and luminance to affect depth perception in handheld mobile AR. Increased fidelity increased perceptual differences among the color and luminance conditions for both shapes, and the simple high-fidelity object was more affected by color and luminance than either of the complex objects.

In the future, we plan to investigate the effects of object shape, fidelity, color, and luminance in a more ecologically valid application context. It would be important to note if these effects are present in a mobile AR application that requires a task, such as an assembly or training task. Depth perception of objects can be extremely important in many AR applications, and it could be advantageous to developers to understand the effects of depth cues on 3D objects.

## ACKNOWLEDGMENTS

The authors would like to thank the reviewers for their valuable suggestions and comments. This work is supported in part by NSF Award IIS-1917728.

## REFERENCES

[1]   H. Adams, "Resolving Cue Conflicts in Augmented Reality," *2020 IEEE Conf. Virtual Real. 3D User Interfaces Abstr. Work.*, pp. 547–548, 2020, doi: 10.1109/VRW50115.2020.00125.

[2]   M. L. Ashley, "Concerning the Significance of Intensity of Light in Visual Estimates of Depth," *Psychol. Rev.*, vol. 5, no. 6, pp. 595–615, 1898.

[3]   R. J. Bailey, C. M. Grimm, and C. Davoli, "The real effect of warm-cool colors." 2006.

[4]   M. Berning, D. Kleinert, T. Riedal, and M. Beigl, "A Study of Depth Perception in Hand-Held Augmented Reality using Autostereoscopic Displays," *Int. Symp. Mix. Augment. Real.*, pp. 93–98, 2014.

[5]   D. Brewster, "Notice of a chromatic stereoscope," *London, Edinburgh, Dublin Philos. Mag. J. Sci.*, vol. 3, no. 15, pp. 31–31, 1852, doi: 10.1080/14786445208646944.

[6]   C. von Castell, H. Hecht, and D. Oberfeld, "Measuring perceived ceiling height in a visual comparison task," *Q. J. Exp. Psychol.*, vol. 70, no. 3, pp. 516–532, 2017, doi: 10.1080/17470218.2015.1136658.

[7]   D. Chatzopoulos, C. Bermejo, Z. Huang, and P. Hui, "Mobile Augmented Reality Survey: From Where We Are to Where We Go," *IEEE Access*, vol. 5, pp. 6917–6950, 2017, doi: 10.1109/ACCESS.2017.2698164.

[8]   J. Coules, "Effect of Photometric Brightness on Judgments of Distance," *J. Exp. Psychol.*, vol. 50, no. 1, pp. 19–25, 1955.

[9]   H. A. David, *The Method of Paired Comparisons*. 1988.

[10]  M. Dengler and W. Nitschke, "Color stereopsis: A model for depth reversals based on border contrast," *Percept. Psychophys.*, vol. 53, no. 2, pp. 150–156, 1993, doi: 10.3758/BF03211725.

[11]  A. Dey, G. Jarvis, C. Sandor, and G. Reitmayr, "Tablet versus Phone : Depth Perception in Handheld Augmented Reality," *2012 IEEE Int. Symp. Mix. Augment. Real. (ISMAR).*, no. November, pp. 187–196, 2012, doi: 10.1109/ismar.2012.6402556.

[12]  A. Dey and C. Sandor, "Lessons learned: Evaluating visualizations for occluded objects in handheld augmented reality," *Int. J. Hum. Comput. Stud.*, vol. 72, no. 10–11, pp. 704–716, 2014, doi: 10.1016/j.ijhcs.2014.04.001.

[13]  B. Dresp, S. Durand, and S. Grossberg, "Depth perception from pairs of overlapping cues in pictorial displays," *Spat. Vis.*, vol. 15, no. 3, pp. 255–276, 2002, doi: 10.1163/15685680260174038.

[14]  P. Dunn-Rankin, G. A. Knezek, S. Wallace, and S. Zhang, *Scaling Methods*. 2004.

[15]  M. Farne, "Brightness as an indicator to distance: relative brightness per se or contrast with the background?," *Perception*, vol. 6, pp. 287–293, 1977.

[16]  D. E. Fox and K. I. Joy, "On polyhedral approximations to a sphere," in *Proceedings. Computer Graphics International (Cat. No.98EX149)*, 1998, pp. 426–432, doi: 10.1109/CGI.1998.694296.

[17]  M. Fujimura and C. Morishita, "Depth Representation Method by Color Tone for 3D Graphics Modeler," *2011 Int. Conf. Complex, Intelligent, Softw. Intensive Syst.*, pp. 639–342, 2011.

[18]  L. Gombač, K. Č. Pucihar, M. Kljun, P. Coulton, and J. Grbac, "3D virtual tracing and depth perception problem in mobile AR," *Conf. Hum. Factors Comput. Syst. - Proc.*, vol. 07-12-May-, pp. 1849–1856, 2016, doi: 10.1145/2851581.2892412.

[19]  C. R. C. Guibal and B. Dresp, "Interaction of color and geometric cues in depth perception: when does 'red' mean 'near'?," *Psychol. Res.*, vol. 69, no. 1–2, pp. 30–40, 2004, doi: 10.1007/s00426-003-0167-0.

[20]  H. Hartridge, "THE VISUAL PERCEPTION OF FINE DETAIL," *Optom. Vis. Sci.*, vol. 25, no. 3, pp. 148–152, 1948, doi: 10.1097/00006324-194803000-00012.

[21]  H. Hartridge, "THE VISUAL PERCEPTION OF FINE DETAIL," *Optom. Vis. Sci.*, vol. 25, no. 3, pp. 148–152, 1948, doi: 10.1097/00006324-194803000-00012.

[22]  A. K. Hebborn, H. Nils, and M. Stefan, "Occlusion Matting : Realistic Occlusion Handling for Augmented Reality Applications," *2017 IEEE Int. Symp. Mix. Augment. Real.*, 2017, doi: 10.1109/ISMAR.2017.23.

[23]  E. H. Johns and F. C. Sumner, "Relation of the Brightness Differences of Colors to their Apparent Distances," *J. Psychol.*, vol. 26, no. 1, pp. 25–29, 1948, doi: 10.1080/00223980.1948.9917393.

[24]  M. Kalia, zu S. C. Berge, H. Roodaki, C. Chakraborty, and N. Navab, "Interactive Depth of Focus for Improved Depth Perception," *Med. Imaging Augment. Real.*, pp. 221–232, 2016, doi: 10.1007/978-3-319-43775-0.

[25]  L. Kjelldahl and M. Prime, "A study on how depth perception is affected by different presentation methods of 3D objects on a 2D display," *Comput. Graph.*, vol. 19, no. 2, pp. 199–202, 1995, doi: 10.1016/0097-8493(94)00143-M.

[26]  E. Kruijff and J. E. S. Ii, "Perceptual Issues in Augmented Reality Revisited," *IEEE Int. Symp. Mix. Augment. Real. 2010*, pp. 3–12, 2010.

[27]  P. Ledda, A. Chalmers, T. Troscianko, and H. Seetzen, "Evaluation of tone mapping operators using a High Dynamic Range display," *ACM Trans. Graph.*, vol. 24, no. 3, pp. 640–648, 2005, doi: 10.1145/1186822.1073242.

[28]  Y. Lee, "Binocular Depth Perception of Stereoscopic 3D Line Drawings," *Proc. ACM Symp. Appl. Percept.*, pp. 31–34, 2013.

[29]  J. M. Liu, G. Narasimham, J. K. Stefanucci, S. Creem-Regehr, and B. Bodenheimer, "Distance Perception in Modern Mobile Augmented Reality," *2020 IEEE Conf. Virtual Real. 3D User Interfaces Abstr. Work.*, pp. 196–200, 2020, doi: 10.1109/VRW50115.2020.00042.

[30]  M. Luckiesh, "On " Retiring " and " Advancing " Colors," *Am. J. Psychol.*, vol. 29, no. 2, pp. 182–186, 1918.

[31]  E. MacCurdy, *The Notebooks of Leonardo Da Vinci*. London, England: Jonathan Cape Ltd., 1938.

[32]  R. K. Mantiuk, A. Tomaszewska, and R. Mantiuk, "Comparison of four subjective methods for image quality assessment," *Comput. Graph. Forum*, vol. 31, no. 8, pp. 2478–2491, 2012, doi: 10.1111/j.1467-8659.2012.03188.x.

[33]  C. N. McCain and C. A. Karr, "Color and Subjective Distance," 1970.

[34]  R. P. O'Shea, S. G. Blackburn, and H. Ono, "Contrast as a depth cue," *Vision Res.*, vol. 34, no. 12, pp. 1595–1604, 1994, doi: 10.1016/0042-6989(94)90116-3.

[35]  N. Olguntürk, "Psychological color effects," *Encyclopedia of color science and technology*. Springer, pp. 1077–1080, 2016.

[36] N. Olguntürk, "Psychological color effects," *Encyclopedia of color science and technology*. Springer, pp. 1077–1080, 2016.
[37] T. Oyama, Y. Tanaka, and Y. Chiba, "Affective dimensions of colors: A cross-cultural study," *Jpn. Psychol. Res.*, vol. 4, no. 2, pp. 78–91, 1962, doi: 10.4992/psycholres1954.4.78.
[38] M. C. Payne, "Color as an independent variable in perceptual research," *Psychol. Bull.*, vol. 61, no. 3, pp. 199–208, 1964, doi: 10.1037/h0046183.
[39] M. C. Payne, "Color as an independent variable in perceptual research," *Psychol. Bull.*, vol. 61, no. 3, pp. 199–208, 1964, doi: 10.1037/h0046183.
[40] M. Pedaste, G. Mitt, and T. Jürivete, "What Is the Effect of Using Mobile Augmented Reality in K12 Inquiry-Based Learning?," *Educ. Sci.*, vol. 10, no. 4, p. 94, 2020.
[41] W. B. Pillsbury and B. R. Schaefer, "A Note on 'Advancing and Retreating' Colors," *Am. J. Psychol.*, vol. 49, no. 1, pp. 126–130, 1937.
[42] W. B. Pillsbury and B. R. Schaefer, "A Note on 'Advancing and Retreating' Colors," *Am. J. Psychol.*, vol. 49, no. 1, pp. 126–130, 1937.
[43] M. Rerabek, P. Hanhart, P. Korshunov, and T. Ebrahimi, "Subjective and Objective Evaluation of HDR Video Compression," in *In 9th International Workshop on Video Processing and Quality Metrics for Consumer Electronics (VPQM)*, 2015.
[44] C. S. Rosales *et al.*, "Distance judgments to on- and off-ground objects in augmented reality," *2019 IEEE Conf. Virtual Real. 3D User Interfaces*, pp. 237–243, 2019, doi: 10.1109/VR.2019.8798095.
[45] D. Schmalstieg and T. Hollerer, *Augmented Reality: Principles and Practice*. 2016.
[46] S. Siegel and N. J. J. Castellan, *Nonparametric Statistics for the Behavioral Sciences*. 1988.
[47] G. Singh, S. R. Ellis, and J. E. Swan, "The Effect of Focal Distance, Age, and Brightness on Near-Field Augmented Reality Depth Matching," *IEEE Trans. Vis. Comput. Graph.*, vol. 26, no. 2, pp. 1385–1398, 2018, doi: 10.1109/TVCG.2018.2869729.
[48] B. B. Smith and M. G. Kendall, "On the Method of Paired Comparisons," *Biometrika*, vol. 31, no. 3/4, pp. 324–345, 1940, doi: doi:10.2307/2332613.
[49] T. H. Starks and H. A. David, "Significance Tests for Paired-Comparison Experiments," *Biometrika*, vol. 48, no. 1/2, pp. 95–108, 1961.
[50] J. M. Sundet, "Effects of colour on perceived depth," *Scand. J. Psychol.*, vol. 19, no. 1, pp. 133–143, 1978, doi: doi:10.1111/j.1467-9450.1978.tb00313.x.
[51] J. E. Swan, A. Jones, E. Kolstad, M. A. Livingston, and H. S. Smallman, "Egocentric depth judgments in optical, see-through augmented reality," *IEEE Trans. Vis. Comput. Graph.*, vol. 13, no. 3, pp. 429–442, 2007, doi: 10.1109/TVCG.2007.1035.
[52] J. E. Swan, L. Kuparinen, S. Rapson, and C. Sandor, "Visually Perceived Distance Judgments: Tablet-Based Augmented Reality Versus the Real World," *Int. J. Hum. Comput. Interact.*, vol. 33, no. 7, pp. 576–591, 2017, doi: 10.1080/10447318.2016.1265783.
[53] J. E. Swan, G. Singh, and S. R. Ellis, "Matching and Reaching Depth Judgments with Real and Augmented Reality Targets," *IEEE Trans. Vis. Comput. Graph.*, vol. 21, no. 11, pp. 1289–1298, 2015, doi: 10.1109/TVCG.2015.2459895.
[54] I. L. Taylor and F. C. Sumner, "Actual Brightness and Distance of Individual Colors when Their Apparent Distance is Held Constant," *J. Psychol.*, vol. 19, pp. 79–85, 1945, doi: 10.1080/00223980.1945.9917222.
[55] W. B. Thompson, P. Willemsen, A. A. Gooch, S. H. Creem-Regehr, J. M. Loomis, and A. C. Beall, "Does the quality of the computer graphics matter when judging distances in visually immersive environments?," *Presence Teleoperators Virtual Environ.*, vol. 13, no. 5, pp. 560–571, 2004, doi: 10.1162/1054746042545292.
[56] L. L. Thurstone, "A Law of Comparative Judgment," *Psychol. Rev.*, vol. 34, no. 4, p. 273, 1927.
[57] E. W., "On the production of shadow and perspective effects by difference of colour," *Brain*, vol. 16, no. 19, pp. 191–202, 1893.
[58] A. S. Yee and P. Milgram, "The use of paired comparisons for evaluating complex route matching performance in a spatial awareness task," *Proc. Hum. Factors Ergon. Soc.*, pp. 1229–1233, 2013, doi: 10.1177/1541931213571273.
[59] M. J. Young, M. S. Landy, and L. T. Maloney, "A perturbation analysis of depth perception from combinations of texture and motion cues," *Vision Res.*, vol. 33, no. 18, pp. 2685–2696, 1993, doi: 10.1016/0042-6989(93)90228-O.